\documentclass[letterpaper, 12pt]{article}
\usepackage[english]{babel}
\usepackage[utf8x]{inputenc}
\usepackage[T1]{fontenc}

\usepackage{amsmath}
\usepackage{amssymb}
\usepackage{graphicx}
\usepackage{natbib}
\usepackage{wasysym} 
\usepackage{subcaption}
\usepackage[colorinlistoftodos]{todonotes}
\usepackage[colorlinks=true, allcolors=red]{hyperref}
\usepackage{lscape}
\usepackage[vlined,ruled]{algorithm2e} 

\usepackage{times}
\usepackage[top=1in,bottom=1in,left=1in,right=1in]{geometry}
\newcommand{\bbeta}{\boldsymbol{\beta}}

\newcommand{\bx}{\boldsymbol{x}}
\newcommand{\bD}{\boldsymbol{D}}

\newcommand{\bY}{\boldsymbol{Y}}

\newcommand{\bB}{\boldsymbol{B}}

\newcommand{\bM}{\boldsymbol{M}}

\newcommand{\bw}{\boldsymbol{w}}
\newcommand{\bd}{\boldsymbol{d}}
\newcommand{\bT}{\boldsymbol{T}}

\newcommand{\bX}{\boldsymbol{X}}

\newcommand{\by}{\boldsymbol{y}}

\newcommand{\bOne}{\boldsymbol{1}}

\title{Enhancing the Demand for Labour survey by including~skills from online job advertisements using model-assisted calibration}
\author{Beręsewicz Maciej\footnote{Corresponding author: maciej.beresewicz@ue.poznan.pl}, Białkowska Greta, Marcinkowski Krzysztof, \\
Maślak Magdalena, Opiela Piotr, Pater Robert\footnote{This work was supported by the Polish Ministry of Science and Higher Education [DIALOG 0127/2016 to B.M. and P.R.]}, Zadroga Katarzyna}

\date{}

\begin{document}
\maketitle


\begin{abstract}
In the article we describe an enhancement to the Demand for Labour (DL) survey conducted by Statistics Poland, which involves  the inclusion of skills obtained from online job advertisements. The main goal is to provide estimates of the demand for skills (competences), which is missing in the DL survey. To achieve this, we apply a data integration approach combining traditional calibration with the LASSO-assisted approach to correct representation error  in the online data. Faced with the lack of access to unit-level data from the DL survey, we use estimated population totals and propose a~bootstrap approach that accounts for the uncertainty of totals reported by Statistics Poland. We show that the calibration estimator assisted with LASSO outperforms traditional calibration in terms of standard errors and reduces representation bias in skills observed in online job ads. Our empirical results show that online data significantly overestimate interpersonal, managerial and self-organization skills while underestimating technical and physical skills. This is mainly due to the under-representation of occupations categorised as Craft and Related Trades Workers and Plant and Machine Operators and Assemblers.
\end{abstract}

Word count: 5 869. 
\clearpage

\section{Introduction}

The process of matching job seekers with job offers is becoming increasingly complicated given technological development and wider access to knowledge. The structural mismatch between labour supply and demand is one of the most challenging problems that need addressing in the labour market. It requires continuous attention of labour market and educational institutions. This mismatch can be determined by various factors. \citet{boudarbat2012} have observed that it depends more on educational characteristics than demographic and socioeconomic factors. Educational mismatch has been widely analysed from the perspective of levels and fields of education, as well as occupations \citep{somers2018}. 

So far, it has been possible to analyse and address such problems as under-education, over-education, occupational and educational mismatch. However, problems of skills (competences) mismatch, to a~large extent, remain unresolved. They encompass the problems of skills shortage, gap and obsolescence \citep{mcguiness2018}. \citet{chevalier2011} shows that the gap in mean salaries between graduates of different fields of study is smaller than the range of salaries of graduates of the same field of study. \citet{harshbein2017} argue that by looking directly at the skill requirements in job offers rather than relying on assumptions about the skills associated with a~particular occupation, it is possible to document the evolution in skill requirements for this occupation over time. Even though the demand for job-related skills can, to some extent, be evaluated by looking at the occupational and educational composition of jobs, it is impossible to make inferences about the demand for transversal skills.

Workers' skills have been measured from a~macroeconomic perspective in a~number of studies, including the OECD's Survey of Adult Skills from the Program of International Assessment of Adult Competences \citep{oecd-mismatch}. However, continuous research on the demand for skills is scarce. Surveys conducted by national statistical institutions (NSI) provide representative data on vacancies across occupational groups or economic sectors. However, they lack detailed information including measures of skills. \citet{pater2018} show different classifications of skills that can be used to measure skills demand, though no international standard has been established. \citet{ilo-mismatch} states that ”In contrast to unemployment, however, which is generally measured according to international standards, a~uniform typology or measurement framework regarding skills mismatch and related issues, such as skills shortages, is lacking”. One potential source of information on skills demand are online job offers placed by employers or entities that work on their behalf.

Big data and the Internet as a~data source have become an important issue in~statistics, particularly in~official statistics. There are a number of~multinational initiatives (e.g. ESSnet on~Big Data; APPOR Task Force on Big Data; European Centre for the Development of Vocational Training, CEDEFOP) that focus on the quality and suitability of~estimates based on new data sources to complement or supplement existing statistical information. For instance, \citet{cedefop-online-jobs, cedefop-online-trends} provides an overview of job vacancies and trends in EU countries. The ESSnet on Big Data included a~work package devoted to job vacancies (WP 1: Web-scraping -- job vacancies). Its aim is to produce statistical estimates of online job vacancies using suitable techniques and specific methodologies. The intention was to explore a~mix of sources including job search sites, job adverts on enterprise websites, and job vacancy data from third party sources. \citet{ess-net-2017,ess-net-2018} was devoted to web-scraping, text mining, classification and comparison with official statistics. The latter was either at the level of statistical units  (companies) or based on NACE and occupancy variables. The project is being continued with three main tasks devoted to 1) methodological framework, 2) statistical output and 3) implementation requirements of prototypes in the relevant statistical production processes at European and national level \citep{wbp-2019}. 

However, before online data can be used for~official statistics, it is crucial to explore potential sources of representation errors \citep{zhang2012topics, reid2017extending}. In this context, \citet{daas2015big, beresewicz2017two, citro2014multiple} discussed coverage, non-response and measurement errors. \citet{Japec2015, Pfeffermann2015, beresewicz2018} address coverage and non-response error, which can lead to significant bias in~big data sources, in~particular if it is non-ignorable. 

Recently, there has been a growing interest in  research on the use of non-probability samples, including big data, together with probability samples. \citet{kim2018combining, yang2018integration, yang2019doubly, yang2019nearest} have developed a rigorous approach to data integration by means of mass imputation using the nearest neighbour approach and double robust estimation; \citet{chen2018, mcconville2017model} proposed using LASSO regression to conduct model-assisted estimation assuming data are missing at random and \citet{chen2019calibrating} extended this approach assuming that only estimated totals are known. \citet{elliott2017inference, Valliant2019, beresewicz2018, buelens2018comparing} give a~general overview of possible approaches to deal with non-probability samples including pseudo-randomization and the model-based approach, while \citet{Japec2015, citro2014multiple, couper2013sky} provide a~general discussion about modern data sources for statistical purposes.

The online job market has been rapidly developing. Thus, online job offers provide interesting research possibilities. There is a~growing body of economic literature on the use of online job offers, with increasing attention paid to skills (see, e.g., \citet{kuhn2004}, \citet{deming2018}, \citet{colombo2019}, and \citet{pater2018}). \citet{harshbein2017}, \citet{marinescu2018}, and \citet{colombo2019} recognize the issue of online data representativeness. They provide comparisons of online data they use to the data from representative surveys. However, none of the articles contains analysis of online data bias correction.


Currently, the DL survey in Poland is used to produce estimates of vacancies by occupation, economic activity sector (NACE) or company size. The goal of our study was to enhance the DL survey by including information about skills obtained from job advertisements. We reused online data collected for the purpose of The Study of Human Capital (HC) 2011-2014 in the~module devoted to job ads. While different approaches to counteract selection bias are found in the literature, we applied pseudo-randomization (calibration) with modern assisting models (LASSO and Adaptive LASSO). We assumed that the selection bias was ignorable given auxiliary variables.

The article has the following structure. Section \ref{sec-data} is devoted to data sources about the Polish vacancy market, including official statistics and selected non-official sources. This section also describes data used in this study. In section \ref{sec-methods} we describe methods of inference based on non-probability data including the bootstrap procedure to estimate variance when only limited population level data is reported. Section \ref{sec-results} presents empirical results for 11 skills obtained from online data. The article ends with conclusions. 


\section{Data}\label{sec-data}

\subsection{The demand for labour survey}

Currently, the DL survey conducted by Statistics Poland \citep{gus-popyt} is the main source of information about job vacancies in Poland. It is designed to obtain information on the satisfied and unsatisfied demand, i.e. the employed (occupied jobs), the vacancies, the newly created jobs, and the liquidated jobs. In 2005 the format of the survey on labour demand was changed in accordance with the Eurostat requirements in order to keep the survey content and methodology uniform across all EU Member States. Since 2007, the survey has been carried out as a~sample survey and covers entities of the national economy employing at least one person.

The DL survey is carried out as a~probability sample survey. The survey sample of 100,000 units is selected separately for units employing more than 9 persons (50,000), and separately for units employing up to 9 persons (50,000). With regard to large and medium-sized units, the sample is stratified by activity (19 NACE\footnote{The Statistical classification of economic activities in the European Community, abbreviated as NACE, is the classification of economic activities in the European Union (EU); the term NACE is derived from the French Nomenclature statistique des activités économiques dans la Communauté européenne. Various NACE versions have been developed since 1970.} sections) and by province (16 NUTS2 regions), resulting in 304 separate subpopulations. Inside each of the subpopulations, the units are sorted in a descending order according to the number of employees. The largest units in each subpopulation that meet the threshold of the number of employees are included in the survey without sampling. Then the sample of the previously determined size is selected from the remaining parts of particular subpopulations. 

As regards small units, employing up to 9 persons, the main purpose of the survey is to obtain results by NACE section. Allocation is carried out between different NACE sections in order to obtain the same expected precision. Within sections, units are stratified by province and then the sample is selected using the stratified, proportional sampling scheme.

Following a significant change made in the Polish classification of occupations in 2011, data collected in the DL survey before 2011 are not comparable to those collected afterwards. Additionally, in 2018 an additional question was included in the survey questionnaire about whether the responding entity placed job offers in district employment offices (DEOs). Each NUTS4 district (Pol. \textit{powiat}) in Poland has its own DEO. 

The survey suffers from non-response, which amounted to 35.2\% in 2011, 36.6\% in 2012, 38.1\% in 2013 and 38.2\% in 2014. Correction for this error involves multiplying sampling weights by the inversion of response rates within particular strata and calibration to meet the known population totals. 

The survey defines the following terms for the measurement of labour demand:

\begin{itemize}
\item \textbf{Vacancies} are positions or jobs unoccupied owing to labour turnover or newly created that simultaneously meet the following three conditions: (1) were actually unoccupied on the survey day, (2) the employer had made efforts to find people willing to take up the job, (3) if adequate candidates were found to occupy the vacancies, the employer would readily take them in.
\item  \textbf{Newly created jobs} are jobs created in the course of organizational changes, expansion or change of business activity, as well as all jobs available in newly established companies.
\end{itemize}

The DL survey provides quarterly estimates about the number of job vacancies by 1) occupation (9 major groups and sub-groups of more detailed occupations denoted by 2-digit codes; with the exclusion of the 10th major group - occupations in the armed forces), 2) NACE, 3) company size (1-9, 10-49 and 50+ employees), 4) ownership type (public, private) and 5) province (16 units). In addition to marginal distributions, estimates of joint distributions of job offers that are published are limited to two-way interactions. Information about precision, measured by relative standard error, is published on an yearly basis only for marginal distributions of auxiliary variables and varies from 2\% to 20\%.


In the study we examined occupation (2-digit codes), NACE and province as potential auxiliary variables to reduce the selection bias in skills described in job offers. We used only estimated totals reported by Statistics Poland as we did not have access to micro-data from the survey. We decided to disregard \textit{Skilled agricultural, forestry and fishery workers} as this group accounts for less than 1\% of all job vacancies.    Table \ref{tab-est-totals} contains information about estimated vacancies for the first quarters of 2011, 2013 and 2014 based on the DL survey. 

\begin{table}[ht]
\centering
\caption{Estimated total number of vacancies at the end of Q1 based on the DL survey} 
\label{tab-est-totals}
\begin{tabular}{rrr}
  \hline
2011 & 2013 & 2014 \\ 
  \hline
71 775 & 42 889 & 52 725 \\ 
   \hline
\end{tabular}
\end{table}

\subsection{Online job advertisements}

\subsubsection{The Study of Human Capital in Poland}

The Study of Human Capital in Poland (HC), a cross-sectional survey to monitor the labour market, was carried out between 2010-2015. The survey was resumed in 2017 but in a~narrower scope. The survey was conducted by the Polish Agency for Enterprise Development (PAED) and the Centre for Evaluation and Analysis of Public Policies at the Jagiellonian University (CEAPP). The survey consisted of four modules: (1) survey of employers, (2) survey of job offers, (3) working-age population survey and (4) representatives of training institutions. The aim was to keep track of the situation in the Polish labour market, monitor the supply and demand for skills as well as the system of education and professional training in Poland in the period 2010-2015. The data from the survey are freely available from the HC survey website (\url{https://bkl.parp.gov.pl}), the methodology of the study is described in \citet{bkl-competences-det} and the data collection procedure is described in the report of  \citet{bkl-raport-2010}. The focus of the survey is limited to the working-age population.

The goal of the HC survey of job offers was to provide characteristics of skills and occupations included in job offers, not produce estimates of these characteristics for all job offers in Poland. 

The statistical unit defined in the job ads module was \textit{a~unique job offer for a~single position, published on a~given day, excluding internships for students and pupils and jobs in foreign countries}. The survey did not distinguish between seasonal, part-time or full-time job offers. This definition differs from the one used in official statistics because it is related to the job description rather than the vacancy. However, we assumed that information included in the job ad can be taken to reflect the job vacancy. 

A mixed mode of data collection was used. Job offers were obtained from a random sample of 160 public employment offices (DEOs; stratified by 16 provinces) in 2010 and the~job search engine www.Careerjet.pl. Job offers had to meet specific requirements for the day of the survey, which was the 4th Monday of March of every year (except for 2010, when the data collection was conducted in September). Therefore, the survey was designed to be comparable between successive years. 

The survey was carried out in three stages. In the case of DEOs, data were collected from the Central Job Offers Database (CBOP), an online service maintained by the Ministry of Family, Labour and Social Policy. According to the reports, the coverage of selected DEOs was insufficient, which is why DEOs' staff was contacted to collect all current job ads. 

Data from the Careerjet.pl website were collected in  a semi-automated manner. Interviewers took a~screenshot of a~displayed job offer or saved the~page as an~html file, and then entered the data using the copy-and-paste method according to a preset format. Each offer was a~separate text file with a corresponding identifier. Then specially prepared software transformed the dataset for the coding process. 

The job offers were coded according to a~categorization key containing a~list of skills, occupations and other features. Each offer was coded independently by two coders. Table \ref{coding-quality} in the Appendix contains information about the coding precision for occupations and NACE sections indicated by the number of digits in a~given code; the higher the number of digits, the more detailed the occupation specification is. Each year, the coding reliability index was calculated based on a~sample of 100 job offers, which represents the total number of codes used and coding consistency of coders. These ratios are presented in Table \ref{sample-size-bkl}.

\begin{table}[ht!]
\centering
\caption{Sample sizes and the coding precision in the DL survey in the online job ads module}
\label{sample-size-bkl}
\begin{tabular}{lrrrrr}
\hline 
Year & 2010 & 2011 & 2012 & 2013 & 2014 \\
Day & 10th Sep & 28th Mar& 26th Mar & 25th Mar & 28th Mar\\
\hline
Initial Sample & 21 195 & 22 243 & 23 366 & 22 795 & 23 452\\
Final sample & 20 009 & 20 634 & 21 594 & 20 081 & 21 456 \\
\hline
DEOs  & 8 198 & 7 018 & 7 253 & 5 614 & 8 542 \\ 
the Internet (Careerjet.pl) & 11 811 & 13 618 & 14 342 & 14 467 & 12 914 \\
\hline
Coding quality & 0.72 & 0.89  & 0.96 & 0.96 & 0.96 \\
\hline
\end{tabular}
\\
Source: \citet{bkl-raport-2010,bkl-raport-2011,bkl-raport-2012,bkl-raport-2013,bkl-raport-2014}.
\end{table}

Verification of offers consisted in removing duplicate ads and those that did not meet the adopted selection criteria. The database did not include offers of low-quality data (where it was impossible to determine the place of work and the recruitment area, as well as offers with insufficient information). The uniqueness of offers from the second survey edition in 2011 was verified at the level of the database, not at the stage of obtaining job offers. Duplicates were distinguished by comparing (1) publication date, (2) source, (3) city, (4) province, (5) job offer reference number (not the ad's ID), (5) company name and (6) occupation. Without access to the raw data, we assumed that publicly available databases contained unique job offers.

Job offers from DEOs selected for the sample were assumed to be valid for the day of the data collection. In the case of Careerjet.pl, first the job offers registered on the day of data collection were downloaded and coded. The target sample size for each year was set at 20000, which included all job offers collected from from DEOs plus as many ads from Careerjet.pl as necessary to reach the target. Table~\ref{sample-size-bkl} presents the initial sample size (before deduplication) and final sample size for each survey year, including the collection date.

The coding precision was lowest in 2010, which is not surprising as this was the first year of the study. The index increases in the subsequent years. Ads from Careerjet.pl accounted for about 60\% of all job offers collected in the survey. 

Results reported in the article are based on the survey of job offers conducted between 2011 and 2013-2014 (three waves). For 2012, the publicly available dataset contained only 1-digit occupations (9 groups, see Table \ref{data-avail}) and it was not possible to obtain the full dataset from the survey administrator. Therefore, we decided to take the following steps regarding the final dataset: 

\begin{itemize}
\item avoid the underrepresentation of skills in job offers from DEOs by focusing only on the data from the Internet (Careerjet.pl),
\item disregard occupations with single digit code (143 records),
\item disregard the 6$^{th}$ occupation category (i.e. skilled agricultural, forestry and fishery workers) because of the small number of job vacancies reported in the DL survey,
\item disregard the following NACE sections: A (Agriculture, Forestry And Fishing), B (Mining And Quarrying), D (Electricity, Gas, Steam And Air Conditioning Supply), E (Water Supply; Sewerage, Waste Management And Remediation Activities), L (Real Estate Activities) for lack of population totals with estimated standard errors.
\end{itemize}

As a~result, the final dataset for the waves in 2011, 2013 and 2014 consisted of a total of 38 100 observations. There were 34 two-digit occupation codes, 16 provinces and 16 NACE \footnote{We collapsed underrepresented NACE sections and occupation codes for job ads and did the same for the DL survey data. See supplementary materials for the whole data processing report.}.

\subsubsection{Skills measured in the study}

The HC survey proposes a  classification of skills for the analysis of the vacancy market. It~was prepared after reviewing various skills classifications used by different international institutions, including: institutions dealing with statistical data (e.g. the Australian Bureau of Statistics), those that develop skills standards (e.g. National Classification of Professional Standards), and enterprises responsible for the development of professional skills (e.g. O*NET. The Occupational Information Network). For more details see \citet[chap. 2]{bkl-competences-det} attached in the Online Supplementary Materials. The survey distinguished the following skills: 

\begin{enumerate}
\item Artistic -- artistic and creative skills, 
\item Availability -- availability to work for the employer, 
\item Cognitive -- finding and analyzing information, drawing conclusions, 
\item Computer -- working with computers and using the Internet, 
\item Interpersonal -- contacts with others, 
\item Managerial -- managerial skills and organization of work, 
\item Mathematical -- performing calculations, 
\item Office -- organization of and conducting office tasks, 
\item Physical -- physical fitness, 
\item Self-organization -- self-organisation, initiative, punctuality, 
\item Technical -- handling, assembling and repairing equipment. 
\end{enumerate}

A detailed description of the skills categories is presented in Table \ref{competences-details}. During the coding process 1 was used if a given skill was included in the job description, 0 = otherwise. There were almost no missing data in variables denoting skills as the lack of a given skill was indicated by 0. 

\begin{table}[ht!]
\centering
\caption{Share of skills included in job offers by data source based on pooled data for 2011, 2013 and 2014} 
\label{comp-sources}
\begin{tabular}{lrrr}
  \hline
Skill & Careerjet.pl & DEOs & Both \\ 
  \hline
Artistic & 15.8 & 2.2 & 11.2 \\ 
  Availability & 21.0 & 2.9 & 14.8 \\ 
  Cognitive & 20.8 & 1.5 & 14.3 \\ 
  Computer & 33.2 & 8.9 & 25.0 \\ 
  Interpersonal & 55.9 & 6.9 & 39.3 \\ 
  Managerial & 29.2 & 2.0 & 20.0 \\ 
  Mathematical & 0.3 & 0.1 & 0.2 \\ 
  Office & 3.8 & 1.5 & 3.0 \\ 
  Physical & 6.0 & 2.0 & 4.7 \\ 
  Self-organization & 59.1 & 7.6 & 41.6 \\ 
  Technical & 4.3 & 5.1 & 4.6 \\ 
   \hline
\end{tabular}
\end{table}

Table \ref{comp-sources} presents the share of given competences included in job offers according to the data source -- Careerjet.pl (the Internet) and DEOs. For example, self-organisation was included in 59.1\% of all job offers on the Internet while only in 7.6\% of offers in DEOs. Spearman correlation coefficient between shares of competences measured in online offers and DEOs was equal to 0.74.

Data from the Internet were much richer in terms of published content in job offers, compared to DEO data. Employers placing job offers online tended to prepare much more detailed descriptions and therefore managed to better specify their requirements. As regards DEOs, the content (and form) of ads was limited by the input format, which allows the employer to enter the sought-after occupation and any preferences regarding education or knowledge of a~foreign language.  

The design of the HC survey did not include imputation of missing data; for instance 254 records had missing values in the occupation and 268 in the province. The highest number of missing values was recorded for NACE (over 22,000), which is mainly due to the lack of information about the company in the ads. The share of missing data varied between the survey waves as presented in Table \ref{missing-data-wawes}. Therefore, we imputed missing data in occupation, NACE and province based on one nearest neighbour with Gower distance and weights assigned to columns that are based on variable importance from random forest. This approach is implemented in the VIM package \citep{vim-package} and was applied to the original dataset.

\subsubsection{Correlation with auxiliary variables}

\begin{table}[ht!]
\centering
\caption{Cramer's V between skills and occupation, NACE section and province, based on the HC survey pooled data for 2011, 2013 and 2014} 
\label{vcramers}
\begin{tabular}{lrrr}
  \hline
  Skill & Occupation (2 digits) & NACE & Province \\ 
  \hline
  Artistic & 0.22 & 0.11 & 0.05 \\ 
  Availability & 0.15 & 0.14 & 0.05 \\ 
  Cognitive & 0.21 & 0.06 & 0.06 \\ 
  Computer & 0.45 & 0.23 & 0.10 \\
  Interpersonal & 0.42 & 0.23 & 0.06 \\ 
  Managerial & 0.34 & 0.15 & 0.04 \\ 
  Mathematical & 0.05 & 0.02 & 0.03 \\ 
  Office & 0.11 & 0.06 & 0.03 \\ 
  Physical & 0.17 & 0.09 & 0.04 \\ 
  Self-organization & 0.34 & 0.19 & 0.04 \\ 
  Technical & 0.31 & 0.11 & 0.07 \\ 
   \hline
\end{tabular}
\end{table}

Given limited access to totals estimated in the DL survey, the correlation of auxiliary variables was assessed only for occupation, NACE section and province. Cramer's V correlation coefficients are presented in Table \ref{vcramers}. The most correlated variable is occupation and the least correlated is province. This is reasonable because skills are specified for occupation rather than for the place of work or the company's type of activity. 


The highest correlations are observed for interpersonal, computer and managerial skills, which suggests that the use of auxiliary variables could reduce selection bias in the case of these skills. The weakest relationship is observed for office, physical and mathematical skills, which means that any correction for selection bias based on these variables is not likely to be effective. 


\section{Methods}\label{sec-methods}

\subsection{Data integration approach}

Enhancing probability survey with online data (i.e. non-probability sample) may be achieved by data integration. Table \ref{tab-data-integration} presents the case when population, sample survey and online data are considered. First three columns denote variables available at unit-level data. $\bX$ denote auxiliary variables such as occupation or NACE sector, target variable(s) denoted by $\bY$ and $\bd$ are weights used for inference based on sample survey. Last two columns contain either known $\bT^{\bX}$ or estimated totals $\widehat{\bT}^{\bX}$ for auxiliary variables $\bX$. Note that we assume $\bX$ are available in all sources, while $\bY$ only for online data. For simplicity, we assume that weights available in sample survey are already corrected for coverage and non-response errors. That is often the case when National Statistical Institutions provide unit-level data with only one set of weights. Note that setting presented in table \ref{tab-data-integration} takes into account case when totals for some domains created by $\bX$ are available (either known from the population data or estimated from sample survey).

\begin{table}[ht!]
    \caption{Basic idea of data integration when variables are available at unit-level or domain-level}
    \label{tab-data-integration}
    \centering
    \begin{tabular}{cccccc}
        \hline
        Data source & $\bX$ & $\bY$ & $\bd$ & ${\bT}^X$ & $\widehat{\bT}^X$\\
        \hline
        Population data & \checkmark & -- & -- & \checkmark & -- \\
        Online data (A) & \checkmark & \checkmark & -- &  --  & -- \\
        Sample survey (B) & \checkmark & -- & \checkmark & -- & \checkmark \\
        \hline
    \end{tabular}
    
\end{table}

The goal of data integration is to estimate some quantity (e.g. mean, total) of target variables $\bY$ present only in online data. \citet{elliott2017inference} summarised possible approaches that consider pseudo-randomization (i.e calibration) or model-based approach. In addition, \cite{Kim_2018} consider mass imputation and double robust estimation that take into account propensity score weighting. 

In the paper we consider pseudo-randomization approach in which pseudo-weights from non-probability sample are calibrated to estimated totals $\widehat{\bT}^X$ or estimated total of $\bY$ based on approach introduced by \citet{wu2001model} and further developed for non-probability samples by \citet{chen2016}. Detailed description is presented in the sections below.

\subsection{Traditional calibration}

Calibration was proposed by \citet{deville1992calibration} and is a~method of searching for so called calibrated weights by minimizing the distance measure between the sampling weights and the new weights, which satisfy certain calibration constraints. As a~consequence, when the new weights are applied to the auxiliary variables in the sample, they reproduce the known population totals of the auxiliary variables exactly. It is also important that the new weights should be as close as possible to sampling weights in the sense of the selected distance measure \citep{sarndal2005estimation}.

Following the notation in \citet{chen2019calibrating}, let us define the online (non-probability) sample as $s_{A,t}$ of size $n_{A,t}$ where $t=1,...,T$ denotes the wave. For simplicity, we drop subscript $t$. This sample contains variables of interest $Y_k$, where $k=1,...,K$.  Further, let $\bd^A_{n_A \times 1}$ be a vector of pseudo-weights that are typically set to $N/n_A$ for all units $i \in s_{A,t}$, where $N$ is the size of the target population. In this approach we assume simple random sampling design for sample $s_A$.

Let $\bD^A$ be a diagonal matrix of pseudo-design weights and $\bw_{n_A \times 1}$ be calibrated weights that minimize an expected distance measure with respect to the design of $A$
\begin{equation}
E_{A} \left[ \sum_{i \in s_A} g(w_i, d_i^A) / q_i \right],
\end{equation}

\noindent under the constraint:

\begin{equation}
\sum_{i \in s_A}  w_i \bx_i^T = \bT^{\bX},
\end{equation}

\noindent where $\bT^{\bX}$ is a~row vector of estimated population totals (e.g. from the reference, external probability sample) of sample calibration variables $\bX$ and $g(w_i, d_i^A)$ is a~differentiable function with respect to $w_i$, strictly convex on an interval containing $d_i^A$ and $g(d_i^A, d_i^A) =0$. The commonly used generalized regression (GREG) estimator uses the $\chi^2$ distance  $g (w_i,d_i^A)  = (w_i - d_i^A)^2/ d_i^A$. For this distance measure:

\begin{equation}
\bw^{\text{GREG}} = \bd^A + \bD^A \bX(\bX^T\bD^A\bX)^{-1}({\bT}^{\bX}-(\bd^A)^T\bX)^T.
\label{greg-weights}
\end{equation}

The estimate of the population mean of outcome $\by_k$ assuming that we have  $k$ target variables is based on calibrated weights:

\begin{equation}
\widehat{\overline{T}}^{\text{GREG}}_{y_k} = \sum_i^{n_A} w^{\text{GREG}}_i y_{ki} / \sum_{i}^{n_A} w_i^{\text{GREG}}.
\label{greg-estim}
\end{equation}

The calibrated weights defined do not rely on any outcome variable. Thus the same set of weights can be applied to all variables in the survey. 

In the case when only estimates of totals $\widehat{\bT}^X$ are known, \citet{dever2010comparison} introduced estimated control calibration. In this framework, we replace ${\bT}^X$ in \eqref{greg-weights} with $\widehat{\bT}^X$, which results in 

\begin{equation}
\bw^{\text{ECGREG}} = \bd^A + \bD^A \bX(\bX^T\bD^A\bX)^{-1}(\widehat{\bT}^{\bX}-(\bd^A)^T\bX)^T,
\label{egreg-weights}
\end{equation}

\noindent and thus the estimated mean is given by

\begin{equation}
\widehat{\overline{T}}^{\text{ECGREG}}_{y_k} = \sum_i^{n_A} w^{\text{ECGREG}}_i y_{ki} / \sum_{i}^{n_A} w_i^{\text{ECGREG}}.
\label{egreg-estim}
\end{equation}

Following \citet{chen2019calibrating} we denote this estimator as ECGREG (Estimated control GREG) to distinguish it from GREG with known population totals.

\subsection{Model-assisted calibration}

Following results obtained by \citet{chen2016, chen2018}, we consider a~model-assisted calibration approach using a~plausible model. Model-assited calibration was proposed by \citet{wu2001model} and further extended by the above mentioned authors. The basic idea of model-assisted calibration is as follows. We build $k$~separate models for each target variable $\by_k$ using the same set of covariates denoted by $\bx_k$: 

\begin{equation}
E_\xi(y_{ki} | \bx_{ki}) = \mu(\bx_{ki}, \bbeta_{k}), \; V_\xi(y_{ki}|\bx_{ki}) = v_{ki}^2\sigma^2,
\label{eq-model-calib}
\end{equation}

\noindent where $\bbeta_k = (\beta_{k1}, ..., \beta_{kp})^T$ and $\sigma$ are unknown superpopulation parameters. $\mu(\bx_{ki}, \bbeta_k)$ is a~known function of $\bx_{ki}$ and $\bbeta_k$, and $v_{ki}$ is a~known function of $\bx_{ki}$ or $\mu(\bx_{ki}, \bbeta_k)$. $E_\xi$ and $V_\xi$ are expectation and variance with respect to the model $\xi$. 

Let $\bB_k$ be the finite population (or census) estimate of $\bbeta_k$ and $\hat{\mu}_{ik} = \mu(\bx_{ki}, \hat{\bB}_k)$, where $ \hat{\bB}_k$ is the sample estimate of $\bB_k$. Then , the model-assisted calibrated weights $\bw$ minimize a~distance measure $E_{A} \left[ \sum_{i \in s_A} g(w_i, d_i^A) / q_i \right]$ under constraints $\sum_{i=1}^n w_i = N$ and $\sum_{i=1}^n w_i\hat{\mu}_{ik} = \sum_{i=1}^N \hat{\mu}_{ik}$.  Under $\chi^2$ distance measure with $q_i = 1$, the model-assisted calibrated weights are:

\begin{equation}
\bw^{MC}_k = \bd^A + \bD^A \bM(\bM^T\bD^A\bM)^{-1}({\bT}^{M}-(\bd^A)^T\bM)^T,
\label{eq-mc-weights}
\end{equation}

\noindent where $\bD^A = diag(\bd^A)$, ${\bT}^{M}=(N, \sum_i^{N} \hat{\mu}_i)$ and $\bM=(\bOne^A, (\hat{\mu}^A)_{i \in s_A})$. Note that in this approach we obtain $K$ sets of weights for each $\by_k$ variable separately. In this setting the population mean is given by 

\begin{equation}
\widehat{\overline{T}}^{\text{MC}}_{y_k} = \sum_i^{n_A} w^{\text{MC}}_{ki} y_{ki} / \sum_{i}^{n_A} w_{ki}^{\text{MC}}.
\label{greg-mc-estim}
\end{equation}

If the totals are estimated from the reference, independent probability sample of size $n_B$, then constraints are $\sum_{i=1}^{n^A} w_i = \sum_{i=1}^{n_B} d_{i}^B$ and $\sum_{i=1}^{n^A} w_i\hat{\mu}_{ik} = \sum_{i=1}^{n_B} d_i^B\hat{\mu}_{ik}$, where $\bd^B$ are weights from probability sample $B$. Similarly, as in the case of GREG, we replace $\bw^{MC}$ with the $\bw^{ECMC}$ obtained from the estimated totals and get

\begin{equation}
\widehat{\overline{T}}^{\text{ECMC}}_{y_k} = \sum_i^{n_A} w^{\text{ECMC}}_{ki} y_{ki} / \sum_{i}^{n_A} w_{ki}^{\text{ECMC}}.
\label{greg-ecmc-estim}
\end{equation}

Further, we assume that $\mu(\cdot)$ is defined as a generalized linear model (i.e. logistic regression), LASSO and adaptive LASSO regression described in the following section.

\subsection{Model-assisted calibration using adaptive LASSO}

Least Angle Shrinkage and Selection Operator (LASSO) is a~regularized regression that can perform both variable selection and parameter estimation (\citet{tibshirani1996regression}); it gained popularity because it prevents model over-fitting by selecting more accurate and parsimonious models. An adaptive LASSO was proposed by \citet{zou2006adaptive}, which in the case of logistic regression assuming $k$ target variables, is given as 

\begin{equation}
\widehat{\beta}_k = \operatorname*{argmin}_{\bbeta_k} 
\left( 
\sum_{i=1}^{n^A} [-y_{ki} (\bx_{ki}\bbeta_{k}) + \log (1 + \exp (\bx_{ki}^T\bbeta_k)] 
+ \lambda_{{n^A}k} \sum_{j=1}^p \alpha_{kj}^{\gamma_k} |\beta_{kj}| 
\right), 
\end{equation}

\noindent where $\alpha_{kj}^{\gamma_k}$ is an adjustable weight and $\gamma_{n^A}$ is a~penalty used to optimize a~model fit measure, while other parameters remain as defined previously. Given $\lambda_{{n^A}k}$ and $\gamma_k$, one can estimate $\widehat{\bbeta}_{k}$ through iterative procedures. Common choice for $\alpha_{kj}$ is $1 / |\widehat{\beta}_{kj}^{\text{MLE}}|$ where $|\widehat{\beta}_{kj}^{\text{MLE}}|$ is the maximum likelihood estimate of $\beta_{kj}$ or $1 / |\widehat{\beta}_{kj}^{\text{RIDGE}}|$  obtained from ridge regression. If $\alpha_{kj}^{\gamma_k}=1$ then we get standard LASSO model. The power of the weight parameter,  $\gamma_k$, is a~constant greater than 0 that interacts with  $kj$ to control LASSO from selecting or excluding parameters. LASSO can be estimated using the glmnet package \citep{glmnet-package}.

Then, to obtain the population mean we need to replace $\bw^{MC}_k$ with corresponding $\bw^{\text{ECLASSO}}_k$ from the standard LASSO model or  $\bw^{\text{ECALASSO}}_k$ obtained under the adaptive LASSO model. To obtain $\widehat{\bbeta}_k$ we followed the approach proposed by \citet{chen2018} and used the cross-validation procedure. For more details refer to \citet{chen2019calibrating}.

\subsection{Estimators used in the paper}

The outcome variable of interest is whether the~description of a job offer ($i=1,...,n_{A,t}$) contained a~given skill. Let us define the binary indicator for the outcome variable $\by_{kt}$ for each $k=1,...,11$-th skill and for each $t = \{2011,2013,2014\}$ 

\begin{equation}
y_{ikt} = 
\begin{cases}
1 & \text{if $i$-th job offer contains $k$-th skill in year $t$}\\
0 & otherwise. 
\end{cases}
\end{equation}

For each variable $k$ we calculate the following estimators: 
\begin{itemize}
\item $\widehat{\overline{T}}^{\text{HTSRS}}_{y_{kt}} = \sum_{i \in s_{A,t}} (N_{A,t}/n_{A,t}) y_{ikt}$, which is Horvitz-Thompson estimator using pseudo-weights,

\item $\widehat{\overline{T}}^{\text{ECGREG}}_{y_{kt}} = \sum_{i \in s_{A,t}} w_{it}^{\text{ECGREG}}y_{ikt} / \sum_{i \in s_{A,t}} w_{it}^{\text{ECGREG}}$, where we use estimated totals for occupation (2-digit code; 34 levels). See Table \ref{tab-occu2-stats}.

\item $\widehat{\overline{T}}^{\text{ECMC}}_{y_{kt}} = \sum_{i \in s_{A,t}} w_{it}^{\text{ECMC}} y_{ikt} / \sum_{i \in s_{A,t}} w_{it}^{\text{ECMC}}$, where we use a logistic regression model for each $y_{k}$ separately based on pooled data from all periods and one auxiliary variable denoting occupation (2-digit code; 34 levels).

\item $\widehat{\overline{T}}^{\text{ECLASSO1}}_{y_{kt}} = \sum_{i \in s_{A,t}} w_{ikt}^{\text{ECLASSO1}}y_{ikt} / \sum_{i \in s_{A,t}} w_{ikt}^{\text{ECLASSO1}}$, where we use LASSO regression for each $y_{k}$ separately based on pooled data from all periods and one auxiliary variable denoting occupation (2-digit code; 34 levels).

\item $\widehat{\overline{T}}^{\text{ECLASSO2}}_{y_{kt}} = \sum_{i \in s_{A,t}} w_{ikt}^{\text{ECLASSO2}}y_{ikt} / \sum_{i \in s_{A,t}} w_{ikt}^{\text{ECLASSO2}}$, where we use LASSO regression for each $y_{k}$ separately based on pooled data from all periods and two auxiliary variable denoting occupation (2-digit code; 34 levels) and NACE (14 levels).

\item $\widehat{\overline{T}}^{\text{ECALASSO1}}_{y_{kt}} = \sum_{i \in s_{A,t}} w_{ikt}^{\text{ECALASSO1}}y_{ikt} / \sum_{i \in s_{A,t}} w_{ikt}^{\text{ECALASSO1}}$, where we use adaptive LASSO regression with the seame settings as ECLASSO1.

\end{itemize}

\subsection{Variance estimation}\label{variance}

\citet{chen2018, chen2019calibrating} proposed analytical formulas for the asymptotic design variance which consists of two parts: 1) variance with respect to non-probability sample $A$, and 2) variance with respect to probability sample $B$. However, this approach requires access to unit-level data from the $s_B$ sample,  which is not always the case. For example, these data cannot be obtained owing to the risk of disclosure or the cost of purchasing these data is very high. 

Moreover, the estimated totals and their uncertainties can only be published in a~limited form. For instance, the DL survey reports standard errors for the estimated totals of  vacancies by size, type of company and NACE section separately. In addition, estimated errors are only published for the last quarter of each year in the annual report. There are no estimates of uncertainty measures for vacancies by occupation; fortunately, there are cross-classification estimates of vacancies for occupation by NACE. Table \ref{tab-rel-var} presents estimated relative standard errors reported by Statistics Poland for the DL survey for 2011, 2013 and 2014. The precision varies between domains defined by NACE section but, in almost all cases, is lower than 20\%. The highest standard errors are for Accommodation and Catering, and Administrative and Support Service Activities, while the lowest -- for Manufacturing and Public Administration and Defence. Also, the estimates and relative standard errors for vacancies within NACE sections are stable over time.

\begin{table}[ht]
\centering
\caption{Relative standard errors of estimators for vacancies in the Demand for Labour survey for Q4 of 2011, 2013 and 2014} 
\label{tab-rel-var}
\begin{tabular}{lrrr}
  \hline
Section & 2011 & 2013 & 2014 \\ 
  \hline
Total & 3.40 & 4.01 & 3.98 \\ 
\hline
  C -  Manufacturing  & 5.50 & 5.27 & 5.64 \\ 
  F -  Construction  & 13.86 & 19.21 & 15.12 \\ 
  G -  Trade; repair of motor vehicles & 13.69 & 15.75 & 16.33 \\ 
  H -  Transportation and storage & 8.07 & 9.93 & 9.17 \\ 
  I -  Accommodation and catering & 15.99 & 20.78 & 18.26 \\ 
  J -  Information and communication & 6.30 & 7.04 & 11.50 \\ 
  K -  Financial and insurance activities & 7.00 & 8.36 & 7.43 \\ 
  M -  Professional, scientific and technical activities & 8.12 & 8.71 & 12.01 \\ 
  N -  Administrative and support service activities & 23.09 & 12.89 & 17.76 \\ 
  O -  Public administration and defence; compulsory social security  & 3.19 & 3.50 & 2.56 \\ 
  P -  Education & 8.85 & 10.65 & 12.06 \\ 
  Q -  Human health and social work activities  & 5.53 & 6.88 & 6.00 \\ 
  R -  Arts, entertainment and recreation & 7.08 & 8.68 & 9.28 \\ 
  S -  Other service activities & 18.09 & 21.29 & 20.77 \\ 
   \hline
\end{tabular}
\end{table}

In view of the limitations of the reporting procedure in the DL survey, we made the following assumption: \textit{standard errors are similar in a given year and we can approximate standard errors from the 1st quarter based on information from the 4th quarter}. Without access to unit-level data, we could not verify the validity of this assumption but as the estimates of vacancies by NACE (and also by occupation) are stable over time this assumption is likely to be valid.

We used the bootstrap method to account for uncertainty in estimating the model based on $s_{A,t}$ and estimated totals from $s_{B,t}$, which is described in Algorithm~\ref{pseudo-code-boot} below (for simplicity we drop subscript $t$ and also assume that the same totals are used for all $K$ variables).

\vspace{0.5cm}

\begin{algorithm}[H]
\SetAlgoLined
 \For{For i in 1:B}{
  \If{Reference sample data $s_B$}{
  Generate one finite population bootstrap based on the procedure\;
  \begin{enumerate}
      \item Generate $\widehat{\bT}^{\text{NACE}*} \sim N(\widehat{\bT}^{\text{NACE}}, \text{SD}(\widehat{\bT}^{\text{NACE}}))$.
      \item Obtain $\widehat{\bT}^{\text{NACE, OCCUP}*} = \widehat{\bT}^{\text{NACE}*} \times \widehat{\bT}^{\text{NACE,OCCUP}} / \widehat{\bT}^{\text{NACE}}$.
      \item Use $\widehat{\bT}^{\text{NACE, OCCUP}*}$ and $\widehat{\bT}^{\text{OCCUP}*}$ as estimated population totals for the estimation procedure.
  \end{enumerate}
  }
  \If{Non-probability sample $s_A$}{
    Generate sample $s_A^*$ from $s_A$ using simple random sampling with replacement 
    }
    For pair $s_A^*, s_B^*$ calculate $\widehat{\overline{T}}^{\text{HTSRS}*}_{y_{k}},  \widehat{\overline{T}}^{\text{ECGREG}*}_{y_{k}},  \widehat{\overline{T}}^{\text{ECMC}*}_{y_{k}},  \widehat{\overline{T}}^{\text{ECLASSO1}*}_{y_{k}},  \widehat{\overline{T}}^{\text{ECLASSO2}*}_{y_{k}},  \widehat{\overline{T}}^{\text{ECALASSO1}*}_{y_{k}}$.
 }
 \caption{Algorithm to obtain bootstrapped population totals from probability sample $B$ and replications of non-probability sample $s_A$}
 \label{pseudo-code-boot}
\end{algorithm}

\vspace{0.5cm}

\noindent where $\text{SD}()$ denotes standard errors derived from Table \ref{tab-rel-var}, $N()$ is normal distribution, $\widehat{\bT}^{\text{NACE, OCCUP}}$ denotes estimated totals for cross-classification of NACE and Occupation (2-digit codes). Note that this part $\widehat{\bT}^{\text{NACE}*} \times \widehat{\bT}^{\text{NACE,OCCUP}} / \widehat{\bT}^{\text{NACE}}$ assumes that we can split $\widehat{\bT}^{\text{NACE}*}$ according to the estimate share of vacancies by occupation in a given NACE section. Table \ref{tab-occup-var} presents information about relative standard errors of the estimated $\widehat{\bT}^{\text{OCCUP}*}$ in the bootstrap procedure. Detailed results are presented in Supplementary materials in Table \ref{tab-occu2-stats}. Uncertainty varies from 2\% to 20\%, which is inline with errors for NACE or other variables reported in the DL survey.

\begin{table}[ht]
\centering
\caption{Relative standard errors of estimators for vacancies by occupation (2-digit code) in Q4 of 2011, 2013 and 2014 based on the proposed bootstrap procedure} 
\label{tab-occup-var}
\begin{tabular}{lrrrrrr}
  \hline
 Year & Min & Q1 & Median & Mean & Q3 & Maximum \\ 
  \hline
  2011 & 2.38 & 4.78 & 6.42 & 7.84 & 10.20 & 20.05 \\ 
  2013 & 3.49 & 4.94 & 6.50 & 8.18 & 10.12 & 20.40 \\ 
  2014 & 2.32 & 5.50 & 6.55 & 8.26 & 10.70 & 17.56 \\ 
   \hline
\end{tabular}
\end{table}

\newpage

The following steps were taken to calculate variance in the bootstrap procedure.  Let $\hat{\theta}^*_{y_{k}} = \widehat{\overline{T}}^{\text{HTSRS}*}_{y_{k}},  \widehat{\overline{T}}^{\text{ECGREG}*}_{y_{k}},  \widehat{\overline{T}}^{\text{ECMC}*}_{y_{k}},  \widehat{\overline{T}}^{\text{ECLASSO1}*}_{y_{k}},  \widehat{\overline{T}}^{\text{ECLASSO2}*}_{y_{k}},  \widehat{\overline{T}}^{\text{ECALASSO1}*}_{y_{k}}$ which are then used to derive variance and relative standard errors (CV) given by the following equations:

\begin{itemize}
    \item Variance
        \begin{equation}
            \operatorname{var}(\hat{\theta}_{y_k})=\frac{1}{B-1} 
            \sum_{b=1}^{B}
            \left(
            \hat{\theta}^*_{y_{ki}}-\overline{\hat{\theta}}^*_{y_{k}}
            \right)^{2}, 
            \quad\; 
            \overline{\hat{\theta}}^*_{y_{k}}= 
            \frac{1}{B} \sum_{b=1}^{B} \hat{\theta}^*_{y_{ki}}
        \end{equation}
    \item Relative Standard Error (CV)
        \begin{equation}
            \operatorname{CV}(\hat{\theta}_{y_k}) = \operatorname{var}(\hat{\theta}_{y_k}) /  \overline{\hat{\theta}}^*_{y_{k}} \times 100\%.
        \end{equation}
\end{itemize}

To estimate variances for the aforementioned estimators we used bootstrap with 500 replicates. We compared the results in terms of relative standard errors. All calculations were done in R statistical software \citep{rcran} using codes written by the authors and LASSO procedure provided in \citet{chen2019calibrating}. Data and R scripts to reproduce all calculations (including estimated models), tables and figures are available at \url{https://github.com/BERENZ/job-offers-bkl} or can be obtained on request.

\section{Estimation of the demand for skills}\label{sec-results}


Table \ref{tab-results-pool} presents point estimates produced by means of the estimators presented in section 3.5. Column HTSRS is used for comparison to verify whether the models corrected the bias resulting from the specificity of online data. All bias-corrected estimates show similar demand for skills.  The biggest differences between the bias-uncorrected (HTSRS) and corrected estimates are visible for the skills with high Cramer's V correlation presented in Table \ref{vcramers}, i.e. interpersonal,  managerial or computer skills.  For almost all categories, online job ads overestimate the share of skills required by employers.

\begin{table}[ht]
\centering
\caption{Point estimates of the fraction of skills for the pooled sample for 2011, 2013 and 2014} 
\small
\label{tab-results-pool}
\begin{tabular}{lrrrrrr}
  \hline
SKILLS & HTSRS & ECGREG & ECMC & ECLASSO1 & ECLASSO2 & ECALASSO1 \\ 
  \hline
Artistic & 15.8 & 12.3 & 12.4 & 12.5 & 13.0 & 12.5 \\ 
  Availability & 20.9 & 19.8 & 19.7 & 19.6 & 21.5 & 19.5 \\ 
  Cognitive & 20.9 & 14.3 & 14.3 & 14.6 & 14.0 & 14.6 \\ 
  Computer & 33.0 & 22.2 & 22.0 & 22.3 & 23.0 & 22.6 \\ 
  Interpersonal & 53.8 & 34.5 & 34.5 & 35.1 & 35.0 & 34.9 \\ 
  Managerial & 26.2 & 16.7 & 16.5 & 16.8 & 17.7 & 16.8 \\ 
  Mathematical & 0.4 & 0.4 & 0.4 & 0.4 & 0.4 & 0.4 \\ 
  Office & 3.9 & 3.1 & 3.1 & 3.2 & 3.4 & 3.2 \\ 
  Physical & 5.4 & 7.4 & 7.6 & 7.5 & 8.2 & 7.6 \\ 
  Self-organization & 58.6 & 43.8 & 43.5 & 43.9 & 46.2 & 43.8 \\ 
  Technical & 4.3 & 7.5 & 7.7 & 7.7 & 8.3 & 7.7 \\ 
   \hline
\end{tabular}
\end{table}

The biggest difference (almost 54\% vs 35\%) between all estimators can be observed for interpersonal competences. Other groups where there is a high difference after adjusting for known population totals are managerial skills (almost 10 p.p.) and computer skills (over 10 p.p.). There are two groups that online jobs underestimate: technical and physical competences. This is mainly due to underrepresentation of two categories of occupations: (7) Craft and related trades workers and (8) Plant and machine operators and assemblers. See Table \ref{tab-occu2-stats} in the supplementary materials.

These results show that the studies not taking into account the extent of selection bias across skill requirements in online job postings, may overvalue or undervalue some skills. For example \citet{deming2018} shows higher relative demand, especially for cognitive and interpersonal (social) skills, and lower demand only for  managerial skills in the US economy. Similarly, \citet{harshbein2017} show higher than ours percent of online postings containing cognitive and computer skills requirements, also for the US economy. However, these differences to a high extent may result from large differences between analysed economies.

\begin{landscape}
\begin{figure}[ht!]
    \centering
    \includegraphics[width=1.5\textwidth]{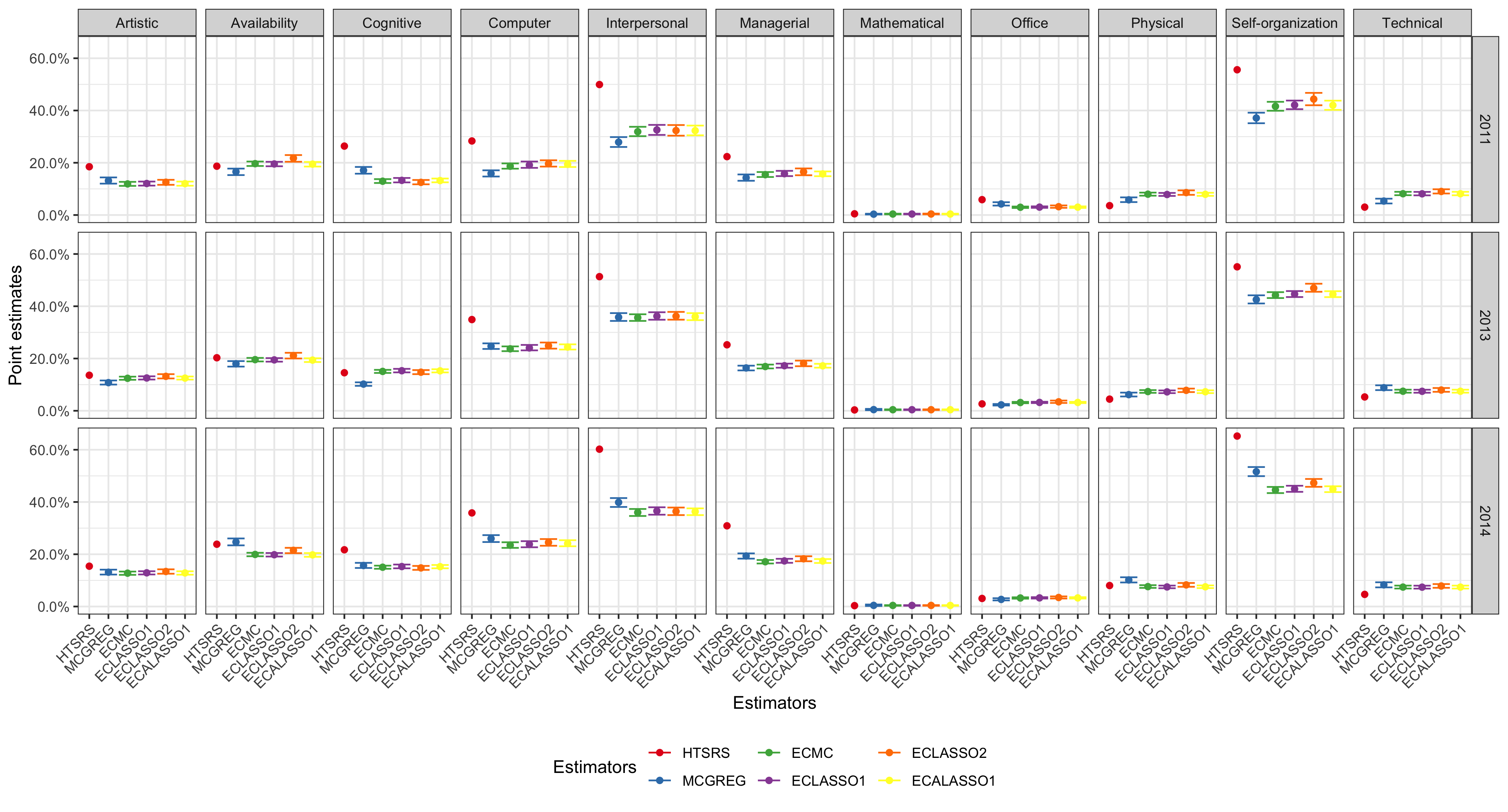}
    \caption{Point estimates for the estimators and HTSRS estimator for each skill for 2011, 2013 and 2014 separately}
    \label{fig-results-wave}
\end{figure}
\end{landscape}

As can be seen, the estimates based on ECGREG, ECLASSO1,2 and ECALASSO1 are similar. This suggests that the variables used for the estimation provide comparable information despite the underlying model.  Table \ref{tab-estim-auc} provides information about Area Under Curve (AUC) for each skill, ECLASSO1, ECLASSO2 and ECALASSO1 model. Based on this table, it can be concluded that the inclusion of two variables -- occupation and NACE -- results in a better model for each skill. The AUC varies from 0.644 for cognitive skills to 0.829 for technical competences, which indicates that the standard LASSO model is better than the adaptive one. Also, there are almost no differences between ECLASSO1  and ECALASSO1, which suggests that despite additional penalty the estimated parameters are close. Figure \ref{fig-results-wave} provides a more detailed comparison of the estimated share of skills over the reference period.

\begin{table}[ht]
\centering
\caption{Average estimates of relative standard errors for skills  over 2011, 2013 and 2014} 
\small
\label{tab-results-pool-se}
\begin{tabular}{lrrrrr}
  \hline
SKILLS & MCGREG & ECMC & ECLASSO1 & ECLASSO2 & ECALASSO1 \\ 
  \hline
Artistic & 11.1 & 3.5 & 3.4 & 3.4 & 3.4 \\ 
  Availability & 22.0 & 1.0 & 0.9 & 1.5 & 1.0 \\ 
  Cognitive & 25.4 & 8.5 & 8.1 & 9.3 & 8.2 \\ 
  Computer & 24.9 & 12.9 & 12.4 & 12.7 & 12.4 \\ 
  Interpersonal & 17.6 & 6.6 & 6.3 & 6.6 & 6.4 \\ 
  Managerial & 15.3 & 5.6 & 5.3 & 5.5 & 5.4 \\ 
  Mathematical & 15.6 & 4.1 & 4.0 & 3.2 & 4.1 \\ 
  Office & 33.5 & 4.7 & 4.4 & 4.4 & 4.6 \\ 
  Physical & 32.6 & 4.1 & 4.2 & 4.7 & 4.3 \\ 
  Self-organization & 16.7 & 3.8 & 3.6 & 3.5 & 3.6 \\ 
  Technical & 25.1 & 5.3 & 5.2 & 7.8 & 5.2 \\ 
   \hline
\end{tabular}
\end{table}

Table \ref{tab-results-pool-se} provides information about estimated relative standard errors for skills estimates for 2011, 2013 and 2014. ECMC and ECLASSO estimators are more efficient than MCGREG and ECMC is less efficient than estimators with LASSO. This is because MCGREG assumes a linear model and auxiliary variables are high dimensional. Note that despite higher AUC for ECLASSO2, it provides less precise estimates mainly due to the  high number of dimensions of the auxiliary variables and variability in totals from the DL survey. Moreover, there are almost no differences between adaptive and non-adaptive LASSO, which suggests that the estimated parameters are probably correctly specified. Based on this result, we can choose estimates based ECLASSO1 as the final ones.

\section{Conclusion}

In the article we described our attempt to enhance the Demand for Labour survey conducted by Statistics Poland by including information about skills listed in online job advertisements. We considered online data as non-probability sample and apply methods that are developed for purpose of integration of probability and non-probability sample. In particular, we applied model-assisted estimators including generalized linear, LASSO and Adaptive LASSO models. Based on these results we conclude that the application of these methods reduced bias in online data for several skills but not for all. This can be explained mainly by the small correlation with the auxiliary variables used. 

To our knowledge this is the first attempt to extend labour market surveys conducted by National Statistical Agencies by data from the Internet. Previous applications were devoted to non-probability samples based on web surveys or opt-in panels. Our approach shows that methods developed for non-probability samples may be applied for modern data sources such as big data. The latter is currently discussed in terms of auxiliary variables for small area estimation, nowcasting of selected indicators or creating new official statistics. However, there are some issues that should be discussed in detail.

The main limitation involved in the use of online data and combining them with existing surveys is the lack of auxiliary variables. For example, occupations or NACE information need to be extracted from the ad description or may not be even provided by employers. On the other hand, official statistics about the demand for labour are based on probability samples with restricted access to unit-level data (which limit possible approaches) or estimated totals for a~certain level or cross-classification (often without uncertainty measures). 

More generally, research on non-probability samples shows that using these data for statistics requires availability of good independent data sources. The main sources are either probabilistic samples or administrative records. Not always official statistics collects data that is required for the data integration purpose. 

Another issue is measurement and unit error. In our study we  associated job advertisement with job vacancy that may not be always the case. We also assumed that description included on job ads may be related to job vacancy occupations deported by Statistics Poland. This should be verified in the future by investigating job descriptions reported by entities in the DL survey. 

Finally, in our study we used online data from 2011-2014 that was already coded and did not require text mining extract occupation or skills. These data may be actually used for preparing training data for machine learning. This is because original descriptions of job advertisements are associated with labels suited for machine learning purposes. However, one should keep in mind that data from the past not necessarily may hold for future job advertisements. 

Despite these problems we conclude that online data combined with official statistics can provide a better picture of competences, education and other requirements made by employers and can be used to monitor changes by interested entities. In the time of decreasing response rates and budget cuts using data that is already "out there on the Internet" is tempting but requires a~attention to its quality and selection of appropriate methods of inference.

\bibliographystyle{chicago}
\bibliography{bibliography}

\clearpage

\appendix
\section{Appendix}

\subsection{Skills measured in the online data}
\begin{table}[ht!]
\footnotesize
\caption{Eleven general skills categories used in the Study of Human  Capital in Poland}
\label{competences-details}
\begin{tabular}{p{2.5cm}p{5.5cm}p{8cm}}
\hline
\textbf{Skill} & \textbf{Behavior dimension} & \textbf{Behavior sub-dimension}\\
\hline
Artistic & artistic and creative skills & -- \\
\hline
Availability & availability & readiness to travel frequently;  flexible working hours (no  fixed slots)\\ 
\hline 
Cognitive & seeking an analysis of information, and drawing conclusions & quick summarising of large volumes of text; logical thinking, analysis of facts; continuous learning of new things \\
\hline
Computer & working with computers and using the Internet  & basic knowledge of MS Office-type package; knowledge of specialist software, ability to write applications and author websites; using the Internet: browsing of websites, handling e-mail\\
\hline
Interpersonal & contacts with other people (with colleagues, clients, people in the care) & cooperation within the group; ease in establishing contacts with colleagues and/or clients; being communicative and sharing ideas clearly; solving conflicts between people \\
\hline
Managerial & managerial skills and organisation of work & assigning tasks to other members of staff; coordination of work of other staff; disciplining other staff  – taking them to task;\\
\hline
Mathematical & performing calculations & performing simple calculations; performing advanced mathematical computations \\
\hline
Office & organisation and conducting office works & -- \\
\hline
Physical & physical  fitness & -- \\
\hline
Self-organization  & self-organisation of work and showing initiative (planning and timely execution of tasks at work, efficiency in pursuing a~goal) & independent making of decisions; entrepreneurship and showing initiative; creativity (being innovative, inventing new solutions); resilience to stress; timely completion of planned actions \\ 
\hline
Technical & technical imagination and handling technical devices & handling technical devices; repairing technical devices \\
\hline
\end{tabular}
\\
\end{table}

\clearpage

\subsection{Details about the online data}

\begin{table}[ht!]
\centering
\caption{Coding precision, measured by the number of job offers with codes of differing accuracy (different number of digits) based on pooled data from 2011, 2013 and 2014 for occupation}
\label{coding-quality}
\begin{tabular}{lr}
  \hline
	Number of digits & Job offers  \\ 
  \hline
  	6 digits & 33 966 \\ 
    5 digits & 2 663\\ 
    4 digits & 715 \\ 
    3 digits & 614 \\ 
    2 digits  & 142 \\ 
    1 digits &   138 \\ 
   \hline
\end{tabular}
\end{table}

\begin{table}[ht!]
\centering
\caption{Information available in published datasets from the Study of Human Capital in Poland despite its quality}
\label{data-avail}
\begin{tabular}{cccccc}
\hline
Variable & 2011 & 2012 & 2013 & 2014 \\
\hline
Occupation (up to 6 digits) 	 & X & -- & X & X \\
Occupation (only 1 digit) 	 & X & X & X & X \\
NACE  (up to 3 digits) 		 & X & X & X & X \\
Industry 					 & X & X & X & X \\
Province 				 & X & X & X & X \\
Subregion 					 & X & X & x & X \\
Education 					 & X & X & X & X \\
Foreign languages			 & X & X & X & X \\
Work experience 			 & X & X & X & X \\
\hline

\end{tabular}
\end{table}

\begin{table}[ht!]
\centering
\caption{Percentage of missing data in selected variables in each wave of the Human Capital in Poland survey} 
\label{missing-data-wawes}
\begin{tabular}{lrrr}
  \hline
Variables  & 2011 & 2013 & 2014 \\ 
  \hline
  Occupation & 0.33 & 0.40 & 0.49 \\ 
  NACE & 6.04 & 56.86 & 41.98 \\ 
  Voivodeship & 1.06 & 0.01 & 0.21 \\ 
   \hline
\end{tabular}
\end{table}

\clearpage

\begin{table}[ht]
\centering
\caption{Distribution of Occupation (ISCO-08 2-digit codes) in Population and HC data (average over 2011, 2013 and 2014)} 
\label{tab-occu2-stats}
\scriptsize
\begin{tabular}{lrrr}
  \hline
Occupation & Population & Online data & CV\\ 
  \hline
11 - Chief executives, senior officials and legislators & 0.49 & 1.68 & 4.50 \\ 
12 - Administrative and commercial managers  & 1.70 & 2.22 & 4.53 \\ 
13 - Production and specialized services managers  & 1.27 & 2.02 & 9.56 \\ 
14 - Hospitality, retail and other services managers  & 0.29 & 2.78 & 11.84 \\ 
21 - Science and engineering professionals  & 4.45 & 4.09 & 4.64 \\ 
22 - Health professionals & 3.33 & 1.47 & 6.45 \\ 
23 -  Teaching professional  & 0.91 & 2.00 & 10.75 \\ 
24 - Business and administration professionals  & 6.73 & 14.65 & 3.74 \\ 
25 - Information and communications technology professionals  & 3.94 & 8.17 & 8.10 \\ 
26 - Legal, social and cultural professionals  & 0.71 & 0.91 & 5.54 \\ 
31 - Science and engineering associate professionals  & 1.51 & 1.50 & 6.03 \\ 
32 - Health associate professionals  & 0.79 & 0.58 & 6.31 \\ 
33 - Business and administration associate professional  & 4.37 & 19.33 & 3.67 \\ 
34 - Legal, social cultural and related associate professionals  & 0.97 & 0.53 & 5.66 \\ 
35 - Information and communications technicians & 1.73 & 0.78 & 6.73 \\ 
41 - General and Keyboard Clerks & 1.41 & 1.82 & 2.90 \\ 
42 - Customer Services Clerks & 5.20 & 2.53 & 6.70 \\ 
43 - Numerical and Material Recording Clerks & 1.28 & 1.46 & 6.63 \\ 
44 - Other Clerical Support Workers & 2.52 & 0.51 & 4.09 \\ 
51 - Personal Services Workers & 2.41 & 3.10 & 16.14 \\ 
52 - Sales Workers & 8.79 & 16.49 & 15.35 \\ 
54 - Protective Services Workers & 1.28 & 1.16 & 13.77 \\ 
71 - Building and Related Trades Workers (excluding Electricians) & 9.53 & 1.71 & 17.09 \\ 
72 - Metal, Machinery and Related Trades Workers & 7.09 & 2.36 & 5.27 \\ 
73 - Handicraft and Printing workers & 0.72 & 0.25 & 5.76 \\ 
74 - Electrical and Electronics Trades Workers & 2.09 & 1.23 & 13.99 \\ 
75 - Food Processing, Woodworking, Garment and Other Craft and Related Trades Workers & 5.64 & 1.18 & 5.35 \\ 
81 - Stationary Plant and Machine Operators & 2.71 & 0.35 & 5.19 \\ 
82 - Assemblers & 2.72 & 0.21 & 6.27 \\ 
83 - Drivers and Mobile Plant Operators & 7.99 & 1.67 & 9.23 \\ 
91 - Cleaners and Helpers & 1.35 & 0.19 & 10.09 \\ 
93 - Labourers in Mining, Construction, Manufacturing and Transport & 2.20 & 0.49 & 8.52 \\ 
94 - Food Preparation Assistants & 1.26 & 0.26 & 19.34 \\ 
96 - Refuse Workers and Other Elementary Workers & 0.60 & 0.31 & 5.41 \\ 
   \hline
\end{tabular}
\end{table}

\clearpage

\subsection{Estimation process and results}

\begin{table}[ht]
\centering
\caption{Quality of the model measured by Area Under Curve (AUC; average over 500 bootstrap replicated)} 
\label{tab-estim-auc}
\begin{tabular}{lrrr}
  \hline
SKILLS & ECLASSO1 & ECLASSO2 & ECALASSO1 \\ 
  \hline
  Technical & 0.829 & 0.846 & 0.829 \\ 
  Mathematical & 0.784 & 0.818 & 0.784 \\ 
  Artistic & 0.665 & 0.672 & 0.665 \\ 
  Computer & 0.748 & 0.755 & 0.748 \\ 
  Cognitive & 0.644 & 0.654 & 0.644 \\ 
  Managerial & 0.722 & 0.731 & 0.722 \\ 
  Interpersonal & 0.731 & 0.750 & 0.731 \\ 
  Self-organization & 0.695 & 0.708 & 0.695 \\ 
  Physical & 0.687 & 0.713 & 0.687 \\ 
  Availability & 0.605 & 0.635 & 0.604 \\ 
  Office & 0.671 & 0.681 & 0.670 \\ 
   \hline
\end{tabular}
\end{table}

\clearpage

\end{document}